\documentclass[12pt]{article}
\usepackage{a4wide}
\usepackage{amssymb}
\begin{document}
{\renewcommand{\thefootnote}{\fnsymbol{footnote}}
%\hfill  IGC--yy/m--n\\
%\medskip
\begin{center}
{\LARGE Comment on ``New variables\\[2mm] for 1+1 dimensional gravity'' }\\
\vspace{1.5em}
Martin Bojowald,\footnote{e-mail address: {\tt bojowald@gravity.psu.edu}}$^1$
Suddhasattwa Brahma\footnote{e-mail address: {\tt suddhasattwa.brahma@gmail.com}}$^{2,1}$
and Juan D.~Reyes\footnote{e-mail address: {\tt jdrp75@gmail.com}}$^3$
\\
\vspace{0.8em}
$^1$ Institute for Gravitation and the Cosmos,\\
The Pennsylvania State
University,\\
104 Davey Lab, University Park, PA 16802, USA\\[2mm]
$^2$ Center for Field Theory and Particle Physics,\\
Fudan University, 200433 Shanghai, China\\[2mm]
$^3$ Facultad de Ingenier\'{\i}a,
Universidad Aut\'onoma de Chihuahua, \\
Nuevo Campus Universitario, Chihuahua 31125, Mexico
\vspace{1.5em}
\end{center}
}

\setcounter{footnote}{0}

\begin{abstract}
  The results reported by Gambini, Pullin and Rastgoo in \cite{NewVar11} are
  special cases of a general treatment of canonical variables for dilaton
  gravity models published earlier in \cite{SphSymmPSM}.
\end{abstract}

\vspace{1cm}

Different sets of canonical variables for $1+1$-dimensional models of gravity
without local physical degrees of freedom have been discussed in
\cite{NewVar11}. These variables are related to connection formulations as
used in classical theories underlying loop quantum gravity. All models of this
form are contained in the class of 2-dimensional dilaton models, or
equivalently in the class of Poisson Sigma models
\cite{Ikeda,PSM,Strobl}. Since these classes had already been formulated in
terms of connection variables \cite{SphSymmPSM}, there should be a strict
relation between the different sets of variables. In this comment we work out
the relationship.

We start with standard formulations of $1+1$-dimensional actions for a dyad
$e^a$ with volume form $\epsilon$, a connection 1-form $\omega$ and a dilaton
field $\phi$. For our purpose here, it suffices to consider torsion-free
models, such that the condition ${\rm D}e^a=0$, using the covariant derivative
given by $\omega$, is implemented by Lagrange multipliers $X_a$.  The dilaton
gravity action with potential $V(\phi)$ is then
\begin{equation}
 S=-\frac{1}{2G} \int_M \left(\phi{\rm d}\omega+ \frac{1}{2}V(\phi) \epsilon+
 X_a{\rm D}e^a\right)
\end{equation}
and takes, after integrating by parts, the form
\begin{equation}
 S=\frac{1}{2G} \int_M \left(e^a\wedge {\rm d}X_a+ \omega {\rm d}\phi-
   X_a\epsilon^a{}_b \omega\wedge e^b- \frac{1}{2}V(\phi)\epsilon\right)\,.
\end{equation}
Here, $M$ is a $1+1$-dimensional manifold with coordinates $(t,x)$.

In Poisson Sigma models, one organizes the variables in new sets
$X^i=(X^-,X^+,\phi)$, $A_i=(e_x^+,e_x^-,\omega_x)$, and
$\Lambda_i=(e_t^+,e_t^-,\omega_t)$. 
A canonical analysis leads to Poisson brackets
\begin{equation}
 \{X^i(x),A_j(y)\} = 2G\delta^i_j\delta(x-y)\,,
\end{equation}
while the $\Lambda_i$ serve as Lagrange multipliers of first-class constraints
\begin{equation} \label{Ci}
 \tilde{C}^i = \frac{1}{2G} \left((X^i)'+ P^{ij}A_j\right)
\end{equation}
with the Poisson tensor \cite{Ikeda,PSM,Strobl}
\begin{equation}
 P=\left(\begin{array}{ccc} 0 & -\frac{1}{2}V(\phi) & -X^-\\ \frac{1}{2}V(\phi)
     & 0 &     X^+\\ X^- & -X^+ & 0\end{array}\right)\,.
\end{equation}

The variables introduced in \cite{SphSymmPSM} are obtained by using the
``absolute values''
\begin{equation}
 X:=\sqrt{X^+X^-} \quad\mbox{ and }\quad e:=\sqrt{e_x^+e_x^-}
\end{equation}
and boost parameters $\alpha$ and $\beta$ in
\begin{equation}
 X^{\pm} = X \exp(\pm\beta) \quad\mbox{ and }\quad e_x^{\pm}=e
 \exp(\pm\alpha)\,.
\end{equation}
They are related to the canonical variables
\begin{equation} \label{Q}
 Q^e=2X\cosh(\alpha-\beta) \quad\mbox{ and }\quad
 Q^{\alpha}=2eX\sinh(\alpha-\beta)
\end{equation}
with
\begin{equation} \label{PoissonQ}
 \{Q^e(x),e(y)\}=\{Q^{\alpha}(x),\alpha(y)\}=\{\phi(x),\omega_x(y)\}=
2G\delta(x-y) \,.
\end{equation}
The inverse transformation is
\begin{equation}
 X^{\pm} = \frac{eQ^e\mp Q^{\alpha}}{2e} \exp(\pm\alpha)\,.
\end{equation}
In canonical variables, the constraints are
\begin{equation} \label{Cpm}
 \tilde{C}^{\pm} = \frac{1}{2G} \left(\left(\frac{eQ^e\mp
       Q^{\alpha}}{2e}\right)' \pm \frac{eQ^e\mp Q^{\alpha}}{2e}
   (\omega_x+\alpha')\pm \frac{1}{2}V(\phi)e\right)\exp(\pm\alpha)
\end{equation}
and
\begin{equation}
 \tilde{C}^3 = \frac{1}{2G} (\phi'+Q^{\alpha})\,.
\end{equation}

For spherically symmetric gravity, corresponding to a specific dilaton
potential $V(\phi)=-2/\sqrt{\phi}$ \cite{Strobl}, one can compare the new
canonical variables to those used in real connection formulations such as
\cite{SphSymm}, denoted as $(K_{\varphi},E^{\varphi};K_x,E^x;\eta,P^{\eta})$
and with Poisson brackets
\begin{equation}
 \{K_x(x),E^x(y)\}=2G\delta(x-y) \quad,\quad
 \{K_{\varphi}(x),E^{\varphi}(y)\}=G\delta(x-y)\,. 
\end{equation}  
(Note that there is no factor of two in the second equation because the pair
$(K_{\varphi}, E^{\varphi})$ represents two angular directions which were
independent before symmetry reduction.)  They are related to
$(Q^e,e;Q^{\alpha},\alpha;\phi,\omega_x)$ by the canonical transformation
\begin{eqnarray}
 Q^e &=& 2\sqrt{2} (E^x)^{1/4} K_{\varphi} \quad,\quad
 e=\frac{E^{\varphi}}{\sqrt{2}(E^x)^{1/4}} \label{Qee}\\
 Q^{\alpha} &=& P^{\eta} \quad,\quad \alpha=-\eta\\
 \omega_x &=& -\left(K_x+\frac{E^{\varphi}}{2E^x}K_{\varphi}-\eta'\right)
 \quad,\quad 
 \phi=E^x\,. \label{omegaphi}
\end{eqnarray}
(Unlike the pair $(K_{\varphi}, E^{\varphi})$, the pair $(e,Q^e)$ is defined
with a factor of two in the Poisson bracket (\ref{PoissonQ}).)  These
relations have been derived in \cite{SphSymmPSM}; see Eq.~(42) in this paper.
Also in \cite{SphSymmPSM}, the spherically symmetric Hamiltonian constraint
has been obtained as
\begin{eqnarray} \label{HDil}
 H[N] &=& C^+[2^{-1/2} N\phi^{1/4}\exp(-\alpha)]- C^-[2^{-1/2}
 N\phi^{1/4}\exp(\alpha)]\nonumber\\
&=& -\frac{1}{2G} \int {\rm d}x N
\left(\frac{K_{\varphi}^2E^{\varphi}}{\sqrt{E^x}}+ 2 \sqrt{E^x}
  K_xK_{\varphi}- \frac{1}{2} E^{\varphi}V(E^x)\right.\\
 &&\qquad\qquad\qquad \left.-
  \frac{((E^x)')^2}{4E^{\varphi}\sqrt{E^x}}+
  \frac{\sqrt{E^x}(E^x)'(E^{\varphi})'}{(E^{\varphi})^2}-
  \frac{\sqrt{E^x}(E^x)''}{E^{\varphi}}\right)\,. \nonumber
\end{eqnarray}

For an arbitrary dilaton potential, this formulation generalizes the
connection formulation of spherically symmetric canonical gravity to arbitrary
$1+1$-dimensional dilaton models. Deriving just this kind of result was the
aim of \cite{NewVar11}. In fact, the definitions of \cite{NewVar11} are
nothing but the results of \cite{SphSymmPSM} with different names chosen for
the new variables.  The expression for $e$ in (\ref{Qee}) is the same as (51)
in \cite{NewVar11}, $Q^e$ in (\ref{Qee}) is (57), and $\omega_x$ in
(\ref{omegaphi}) is (60).

There is a different formulation in \cite{NewVar11} for the specific case of
the CGHS model (constant dilaton potential).  This model, as presented in
\cite{NewVar11} has a Hamiltonian constraint with only $K_xK_{\varphi}$ in its
kinetic part but no contribution of $K_{\varphi}^2$, unlike what we have in
(\ref{HDil}). However, this formulation is not new either, even though it does
not correspond to a connection formulation as defined in \cite{SphSymmPSM}. It
rather amounts to using the original canonical variables (\ref{Q}) of dilaton
models, along with $\omega_x$, and just renaming them as $\omega_x\equiv K_x$,
$Q^e\equiv K_{\varphi}$ and $e\equiv E^{\varphi}$. Except for the new names,
these variables, in particular $Q^e$, have been introduced in
\cite{SphSymmPSM}; see Eq.~(20) in this paper. Except for using the
SU(1,1)-invariant $Q^e$, they correspond to a first-order formulation of
Poisson Sigma models. The constraints (\ref{Cpm}) in these variables depend on
$\omega_x$ in a linear fashion, and so does the Hamiltonian constraint
obtained by a linear combination. In these variables, therefore, the
Hamiltonian constraint has only one term $Q^e\omega_x=K_xK_{\varphi}$ but not
contribution from $K_{\varphi}^2$ (which would amount to $(Q^e)^2$, but there
is no such term in (\ref{Cpm})). The quadratic term is introduced in
(\ref{HDil}) by applying the canonical transformation (\ref{Qee}) and
(\ref{omegaphi}) via the product $Q^e\omega_x$ in (\ref{Cpm}).

%We conclude that \cite{NewVar11} did not contain significant new results at
%the time when it was published. 
Even though \cite{NewVar11} cites
\cite{SphSymmPSM}, it misses the close relationship between the results.

\section*{Acknowledgements}

This work was supported in part by NSF grant PHY-1607414.

%\bibliographystyle{../preprint.bst}
%\bibliography{../Bib/QuantGra}

\end{document}